# Resistance standards with calculable, nearly negligible AC-DC difference at frequencies up to 2 MHz for the calibration of precision LCR meters

Jürgen Schurr[1], Rolf H. Judaschke[1] and Shakil A. Awan[2]

[1] Physikalisch-Technische Bundesanstalt (PTB), Bundesallee 100, 38116 Braunschweig, Germany
[2] Nanomaterials and Devices Laboratory, School of Engineering, Computing and Mathematics,
   University of Plymouth, Devon PL4 8AA, United Kingdom

E-mail: Juergen.Schurr@ptb.de, Rolf.Judaschke@ptb.de and shakil.awan@plymouth.ac.uk

**Abstract**

We have developed novel impedance standards based on thin-film surface-mount-device (SMD) resistors. Due to the small dimensions of such resistors, the quantities determining their frequency dependence are very small and can be either measured or numerically calculated. A series connection of thin-film SMD resistors allows us to further improve the DC and the AC properties. The nominal resistance value of our application is 12.906 kΩ but other values are just as possible. At frequencies up to 2 MHz, the calculated frequency dependence amounts to only a few parts per million of the DC value, which is about four orders of magnitude smaller than for all conventional calculable AC-DC resistors having a similar nominal DC value. To measure the frequency dependence, we use a precision inductance-capacitance-resistance (LCR) meter at frequencies up to 2 MHz that has a reproducibility of a few parts per million but a systematic uncertainty which is specified by the manufacturer to increase from 300 parts per million in the lower frequency range to 3000 parts per million at 2 MHz. Measurements of two very different SMD-based resistance standards allow verification of the model calculation as well as the calibration of the precision LCR meter, both with a relative uncertainty of a few parts per million in the whole frequency range. This boost in precision enables new applications in this frequency range such as the verification of conventional calculable resistance standards, the calibration of impedance standards, and future measurements of the quantum Hall resistance.

## 1. Introduction

The past years have brought an increased interest in impedance standards for frequencies from the audio range up to 2 MHz or even 100 MHz [1–8]. The measuring instruments in this frequency range are commercial instruments like precision LCR meters, impedance analysers, and vector network analysers as well as manual coaxial high-frequency bridges developed by national metrology institutes [6–13]. The excellent features of these instruments have led to the development of improved high-frequency techniques and standards [1–8], but further improved standards are required to enable the precise calibration of these instruments in their entire frequency range and to tap their full potential. This in turn will lead to an increasing need for international intercomparisons [14].

Conventional wire resistors with a calculable AC-DC difference (like Haddad [15] and Gibbings [16] resistors) were developed for the audio frequency range. At a nominal resistance value of 12.9 kΩ as needed for this work, the wire must be very long and thin (for example, 0.8 to 3.2 m long at a diameter of 10 to 20 μm). The AC-DC difference increases with the second power of both the frequency and the wire length. As a result, the AC-DC difference at a frequency of a few megahertz is very large and at present cannot be calculated with the desired uncertainty. Moreover, these standards are highly sensitive to vibration and mechanical shock and, thus, not transportable.

To design a resistance standard which is more suitable for high frequencies, we refer to the general rule that the higher the maximum operating frequency is, the smaller the length and the cross-sectional area of the wire must be. Fortunately, these requirements are fulfilled by commercial thin-film SMD resistors. Such resistors consist of a cuboid of ceramic (often $Al_2O_3$) that carries a thin film of a high-resistive alloy like nickel-chrome. SMD resistors are commercially available with a remarkable precision and a broad range of resistance values. Due to their small dimensions, single thin-film resistors at nominal values up to about 100 Ω have good AC and DC properties, and are used in metrology, for example, as shunt resistors [17]. At larger nominal resistance values as desired for this work, the AC properties of a single thin-film resistor are not sufficient for our application. A strong improvement can however be achieved by means of a series connection of several resistors. In any case, such standards are compact, mechanically robust, have a small temperature and power dependence, and can be transported to customers, calibration laboratories, and participants of international intercomparisons.

This work is motivated by developing a resistance standard that enables the calibration of precision LCR meters, here a Keysight E4980A[*]. The resolution of the LCR meter measuring a resistance of 12.9 kΩ at a single frequency is 0.01 Ω, though this is ten times larger at an automatically executed sequence of selectable frequencies (a so-called frequency sweep) as used in this work (i.e., 0.1 Ω corresponding to 7.8 parts per million). The relative uncertainty of this instrument as specified by the manufacturer is a combination of its base uncertainty of $5 \times 10^{-4}$ and of additional, similarly large contributions which depend on the impedance range, the measuring voltage, and the frequency. However, the reproducibility of the instrument over a period of a few weeks is much better, namely a few parts per million at resistances of the order of 10 kΩ [18]. Therefore, a calibration of the LCR meter with an uncertainty of a few parts per million in the whole frequency range by a suitable resistance standard is feasible, at least in the respective resistance range. This will enable new applications of impedance metrology such as measurements of the quantum Hall resistance at frequencies up to 2 MHz, which is what motivates this work. Therefore, the desired resistance value is $R_K/2 = 12906.40373$ Ω corresponding to the second, most commonly used quantum Hall plateau (with $R_K$ the von Klitzing constant). Of course, any other resistance value can be realized by the method presented here.

This paper is organized as follows: in section 2, the model of the series-connected SMD resistors is discussed; section 3 discusses the design of the four-terminal-pair (4TP) resistance standards; finally, section 4 presents the experimental results and the uncertainty achieved.

## 2. Model

### 2.1 Model equation

Thin-film SMD resistors consist of a small cuboid of ceramic which carries a resistive metal film, a protective lacquer layer, and two metal end caps for contacting. The impedance of $N$ series-connected SMD resistors can be calculated using the equivalent circuit of a single SMD resistor shown in figure 1. The relevant quantities are the total resistance $R$, the series inductance $L_S$, the parallel capacitance $C_P$, the capacitance between each end cap and the coaxial case $C_\text{cap}$, the capacitance between the metal film and the coaxial case $C_\text{R}$, and the associated dissipation factors $D_\text{P}$, $D_\text{cap}$, and $D_\text{R}$. Note that the equivalent circuit in figure 1 considers half the resistance and inductance on either side of $C_\text{R}$ to avoid orientation-dependent artifacts of the calculation. By means of the star-delta transformation, the impedance is successively calculated for $N$ = 1, 2, 3, … until the results can be generalised:

$$Z(\omega) = R \left(1 - \omega R \frac{C_\text{P}}{N}[\text{j} + D_\text{P}] + \text{j}\frac{N\omega L_\text{S}}{R} \right.$$
$$+ \omega R C_\text{cap} 2\left\{\frac{N^2-1}{6N}\right\}[\text{j} + D_\text{cap}] + \omega R C_\text{R}\frac{N^2+\frac{1}{2}}{6N}[\text{j} + D_\text{R}]$$
$$- \{\omega R/N\}^2[C_\text{P}^2 - C_\text{P}C_\text{R}/2]$$
$$\left. + \omega^2 L_\text{S}\left\{2C_\text{P} - C_\text{R}\frac{N^2 + 1/2}{3} - C_\text{cap}2\frac{N^2-1}{3}\right\} + \Delta_\text{skin}\right) \quad (1)$$

with $\omega$ the angular frequency. Terms proportional to the third or higher power of the frequency are neglected. The ac resistance $R$ is not to be mistaken for the real part of $Z(\omega)$. As will be discussed in the next section, when the skin effect $\Delta_\text{skin}$ is taken into account, every $R$ in equation (1) is practically equal to the dc resistance. $C_\text{cap}$ could be expressed by the total capacitance to the case, $C_\text{C} = 2C_\text{cap} + C_\text{R}$, because it is easier to measure, but for the sake of simplicity, the resulting equation is not given here. The total equivalent parallel capacitance, $C_\text{P,tot}$, follows from the imaginary part of equation (1):

$$C_\text{P,tot} = \left\{\frac{C_\text{P}}{N} - \frac{NL_\text{S}}{R^2}\right\} - C_\text{cap}2\frac{N^2-1}{6N} - C_\text{R}\frac{N^2+1/2}{6N}. \quad (2)$$

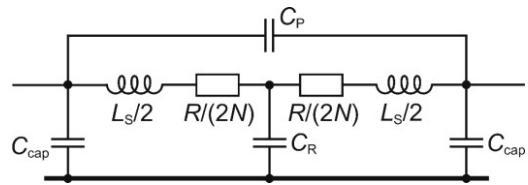

**Figure 1.** Equivalent circuit of a single SMD resistor, with $R$ the total resistance value of $N$ nominally equal SMD resistors to be connected in series. $C_{cap}$ is the capacitance between each metal end cap (plus a connecting wire if applicable) and the coaxial case (bold line).

In equation (1), the dissipation factor of the capacitances give rise to an approximately linear frequency dependence (see also [20]). Firstly, we point out that with increasing $N$, the contribution of the parallel capacitance decreases, whereas the contribution of the case capacitance increases. Secondly, conventional calculable resistance standards often do not consider a possible dissipation factor of the capacitances involved. This might be correct in many cases, but it may also explain the finding that some calculable resistance standards show an unexplained linear frequency dependence [19]. In our case, the relevant dissipation factors are included in the model equation (1) and their quantitative values are measured or at least an upper limit is determined from a measurement. This approach is sufficient here because the corresponding contribution to the frequency dependence will turn out to be smaller than the resolution of the precision LCR meter.

The series inductance of the single SMD resistors in equations (1) and (2) has two effects. Firstly, it corresponds to a negative contribution to the effective parallel capacitance. This contribution will turn out to be practically negligible. As far as we know, the series inductance is not accompanied by any significant hysteretic dissipation. This simplifies the equation and the analysis (otherwise it could be considered as an effective dissipation factor of an effective parallel capacitance without losing the validity of any conclusion of this work). Secondly, the series inductance gives rise to a small quadratic frequency dependence.

Furthermore, equation (1) includes resistive quadratic frequency terms. These terms dominate in the case of a single thin-film SMD resistor as well as calculable wire resistors [15, 16] but are reduced here by a factor $N^2$, which is a great advantage. Thus, both the inductive and the resistive quadratic frequency dependence is very small.

## 2.2 Skin effect

The main reason that the resistance $R$ in equation (1) differs from the DC resistance is the skin effect. It is described by the skin depth $\delta = (2/[\mu_0\omega\sigma])^{1/2}$ with $\mu_0$ denoting the vacuum permeability and $\sigma$ the electrical conductivity of the metal film. To calculate the skin effect on the resistance, it is necessary to know that the thin films used in this work are not plain strips but have a coarse meandering structure (which will be shown below). Imagine that the meander could be unfolded to a straight and uniform metal strip with a rectangular cross section of width $W$ and thickness $H$. In the whole frequency range up to 2 MHz, $W$ and $H$ are much smaller than the skin depth δ. In this approximation, the skin effect on the resistance can be analytically calculated according to [20]:

$$R(\omega) = R_\text{DC}(1 + \Delta_\text{skin}) \text{ with } \Delta_\text{skin} = \frac{1}{12}\frac{HW}{\delta^2} + \frac{1}{12}\frac{H^2}{\delta^2}. \quad (3)$$

The two contributions to $\Delta_\text{skin}$ are due to a weak current redistribution across the width and across the thickness of the metal strip, respectively (compared to a uniform current distribution at DC ). Note that the skin effect on a thin-film resistor in general has a linear frequency dependence because the cross section is essentially one-dimensional (whereas the skin effect of conductors with a circular cross section exhibits a quadratic frequency dependence).

We know neither the electrical conductivity nor the thickness of the metal films but the film thickness is typically in the range of (10 to 100) nm. As this is small compared to the skin depth, the second term of $\Delta_\text{skin}$ in equation (3) is smaller than $2 \times 10^{-9}$ MHz$^{-1}$ and, thus, neglected in the following. In contrast, the first term of $\Delta_\text{skin}$ in equation (3) requires more attention. Interestingly, the DC resistance $R_\text{DC} = l/(\sigma HW)$ with $l$ being the length of the straight metal strip depends on the conductivity $\sigma$, the width $W$, and the height $H$ in the same way as the first term of $\Delta_\text{skin}$. Therefore, the skin effect can be expressed solely by $R_\text{DC}$ and $l$:

$$\Delta_\text{skin} = \frac{1}{24}\frac{\mu_0 \omega l}{R_\text{DC}}. \quad (4)$$

This means that $H$ and $\sigma$ do not need to be known, which saves a lot of effort. The contribution of the metal caps and the soldering joints to the skin effect can be neglected. Finally, every $R$ in equation (1) has to be replaced by $R_\text{DC}(1 + \Delta_\text{skin})$. Because $\Delta_\text{skin}$ is very small, it is sufficient to consider only the first order term in equation (1), as already done. Then, every $R$ in equation (1) is practically equal to the DC resistance.

## 3. Resistance standards

### 3.1 Single thin-film SMD resistors

Thin-film SMD resistors are available from several manufacturers, in a wide range of resistance values, and with different specifications for the temperature coefficient and the deviation from the nominal resistance value. The SMD resistors favoured for this work were either not deliverable, or the delivery time was too long for this project. As a compromise, we chose two different, easily available types described in the following.

The first type is the RG series from Susumu* at the size 0805 (imperial system) with nominal values of 3.0 kΩ and 3.3 kΩ, specified with a tolerance of ±0.1% and a temperature coefficient of ±10 × 10$^{-6}$ per °C. The DC resistance of ten single pieces of each nominal value and from the same batch was measured. Very fortunately, all actual deviations from nominal were in a narrow interval around 0.05%. Thus, we built two series-connected resistors, each consisting of one 3.0 kΩ resistor and three 3.3 kΩ resistors, with a total DC resistance value of about 12.906 kΩ, and a relative deviation from the nominal value $R_K/2$ of only about 10 × 10$^{-6}$.

The second type of thin-film SMD resistor is from Vishay*. We chose resistors at a nominal value of 1.3 kΩ from the TNPW series and 1.0 kΩ from the TNPU series, both with the size 0603 (imperial system) and a resistance tolerance of ±0.05%. The temperature coefficients are specified to be in the range of ±25 × 10$^{-6}$ per °C and ±2 × 10$^{-6}$ per °C, respectively. Combining three 1.3 kΩ resistors and nine 1.0 kΩ resistors gives a total nominal value of 12.900 kΩ. Selecting SMD resistors with the largest DC resistance value from a batch of 50 individually measured pieces of each nominal value allowed us to build two series-connected resistors with a relative deviation at DC from the desired nominal value $R_K/2$ of about -250 × 10$^{-6}$. The combined temperature coefficient is expected to be in the range of ±8 × 10$^{-6}$ per °C, very similar to that of the first quoted type.

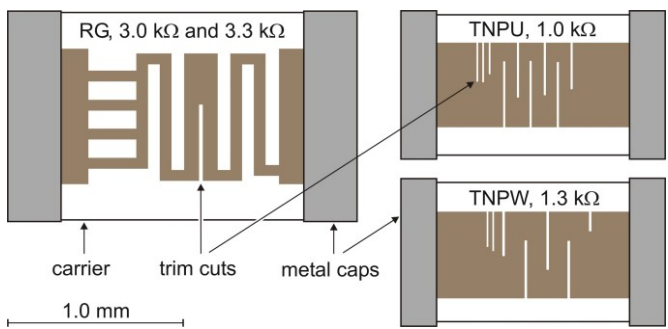

**Figure 2.** Schematic true-to-scale top view of the thin-film SMD resistors used in this work, after removal of the protective lacquer to make the metal film (shown in brown) of the Susumu* SMDs (left) and the Vishay* SMDs (right) visible.

These thin-film SMD resistors have a coarse meandering structure fabricated by the manufacturers with different techniques. To make the meanders visible, we removed the protective lacquer from test resistors by means of a diamond lapping disk. To enable the measurement of the in-plane dimensions of the meander, photomicrographs were made and converted to the schematics shown in figure 2. The Susumu* resistors at the two different nominal values use the same pattern whereas the two Vishay*-type resistors are made with different film technologies and have a different meander.

### 3.2 Numerical calculations

To numerically calculate the series inductance, which could not be determined experimentally, and to verify the quite delicate measurement of the parallel capacitance, a three-dimensional simulation of the thin-films resistors was carried out using the electromagnetic analysis software CST Microwave Studio®*. For each type of thin-film resistor, the ceramic carrier with the resistive meander and the metal caps was discretized into a fine three-dimensional network of meshes at which Maxwell's equations have been solved. From this, the impedance of a single thin film with respect to the soldering caps is determined as a function of frequency. In the simulation, the resistor is considered to float in free space, which means that it has no capacitance to the surroundings. This situation corresponds to the special case of the model equation (1) with $N = 1$, $C_{cap} = 0$, and $C_R = 0$. The resulting model equation with $C_P$ and $L_S$ as free parameters is fitted to the numerically simulated impedance. To allow separation of the inductive quadratic frequency dependence from the otherwise dominating resistive quadratic frequency dependence (see equation (1)), the height of the simulated thin film is temporarily set to an increased value.

The results of the series inductance are 3.3 nH for the Susumu* resistors and 1.6 nH for the Vishay* resistors. They agree reasonably with the analytical calculation of a straight metal strip but are more realistic. As expected, the series inductance is much larger than the 0.1 nH of an unstructured plain metal film but about four orders of magnitude smaller than for metal-foil resistors with an extremely fine meander [21]. This is an invaluable advantage for our work.

### 3.3 Design of resistance standards

The arrangement of the four series-connected SMD resistors is shown in the upper part of figure 3 and discussed in the following. The SMD resistors are mounted free-standing to avoid nearby dielectrics whose permittivity would increase the relevant capacitances and whose dissipation factor could increase the frequency dependence of the resistance standards. Soldering the free-standing SMD resistors directly together could cause mechanical stress. This in turn could give rise to a reduced stability of the resistance and to microcracks in the metal films which may cause an unwanted increase of the resistance and erratic jumps. To avoid such effects, only pairs of two SMD resistors were directly soldered together, and the pairs were connected via flexible arcs of silver-coated copper wire (diameter 0.6 mm) to absorb mechanical stress (as shown at the top of figure 3). The series inductance of the connecting wires and the wire arcs are larger than the series inductance of the thin films, but this is acceptable since the frequency dependence is not dominated by the inductive contribution. The overall length is 15 mm for the four series-connected Susumu* resistors and 31 mm for the twelve series-connected Vishay* resistors.

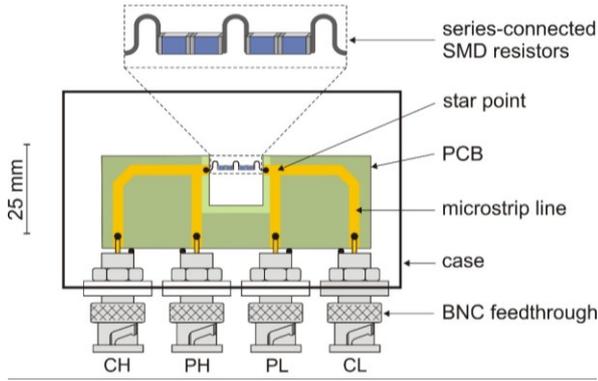

**Figure 3.** Schematic diagram of a 4TP resistance standard based on four series-connected thin-film resistors shown enlarged at the top. The film-protecting lacquer is marked in blue. The PCB with and without a ground plane behind is shown in grey-green and light green, respectively.

The wire arcs and the use of different nominal DC resistance values at each standard break the symmetry of $N$ nominally equal resistors as assumed in equation (1). Instead of calculating a more complicated model equation, we prefer to keep the model simple. We put in the values of capacitance and inductance, respectively, averaged over SMD resistors with different nominal values and with and without wire arcs. Then we estimate an uncertainty. This approach is sufficient here because the calculated frequency dependence is close to the resolution of the LCR meter.

The connections between the series-connected SMD resistors and the four Bayonet Nut Connectors (BNCs) of the case are crucial to success. The higher the maximum operating frequency and the longer the connecting lines, the more important is a uniform characteristic impedance from the BNC ports of the LCR meter to the star points at the series-connected thin-film resistors. Any coarse mismatch would lead to unwanted reflections of the electromagnetic waves and is difficult to be considered accurately. Furthermore, the current and potential lines at the star points should be orthogonal to each other to keep the mutual inductance small. It is also important to keep unwanted stray capacitances to the thin-film resistors small. Less important than these aspects is the routing of the return current in the outer conductor close to the inner conductor, but this helps to keep electrical interference small [1, 6]. In principle, the connections inside the case could be made by miniature coaxial cables [1, 6], but the practical implementation was not satisfying. Instead, we used a printed circuit board (PCB) carrying microstrip lines as shown in figure 3.

A central shield between the PH and the PL line is not necessary. This is because the ground plane on the rear side of the PCB reduces the unwanted stray capacitance between the PH and the PL line to 0.4 fF in the area underneath the cutout and to 0.8 fF in the cutout. As these stray capacitances are so small, the requirements for the dissipation factor of the PCB material are already met with standard FR-4 (i.e., a special low-loss PCB material is not necessary).

As already mentioned above, the characteristic impedance of the microstrip lines on the PCB must match the impedance of the BNC connectors at the resistance standard and the LCR meter. The characteristic impedance can be determined from the measured lead parameters according to $Z_0 = (Z_m/Y_m)^{1/2}$ with $Z_m$ and $Y_m$ being the impedance and the admittance of the microstrip lines or the BNC connectors, respectively. The commercial BNC connectors are designed for a characteristic impedance of 50 Ω at frequencies in the microwave range, but at lower frequencies, the characteristic impedance is substantially larger and decreases with frequency as shown in figure 4 (top). To determine the optimum line width of the microstrips, we fabricated PCBs carrying microstrip lines of different width and measured their characteristic impedance as a function of frequency (figure 4 top). In contrast to BNC connectors, the characteristic impedance of microstrip lines slightly increases with frequency. In both cases, the frequency dependence is attributed to the skin effect that reduces the effective cross section of an electrical conductor with increasing frequency. While smaller cross-sections reduce the inductance of a coaxial cable or connector, it increases the inductance of a microstrip line. This explains the different frequency dependence of the characteristic impedance of microstrip lines and BNC connectors.

A perfect impedance match over the whole frequency range is therefore not possible but also not necessary: we soldered one and the same resistor successively to two PCBs carrying microstrip lines of different width and measured the effect on the frequency dependence of that resistor. From this, the effect of the impedance mismatch is scaled for each line width (figure 4 bottom). Then we chose 2.25 mm as the most suitable line width. The associated effect of the impedance mismatch over the whole frequency range does not exceed $3 \times 10^{-6}$, which is smaller than the resolution of the LCR meter at a frequency sweep and, thus, tolerable.

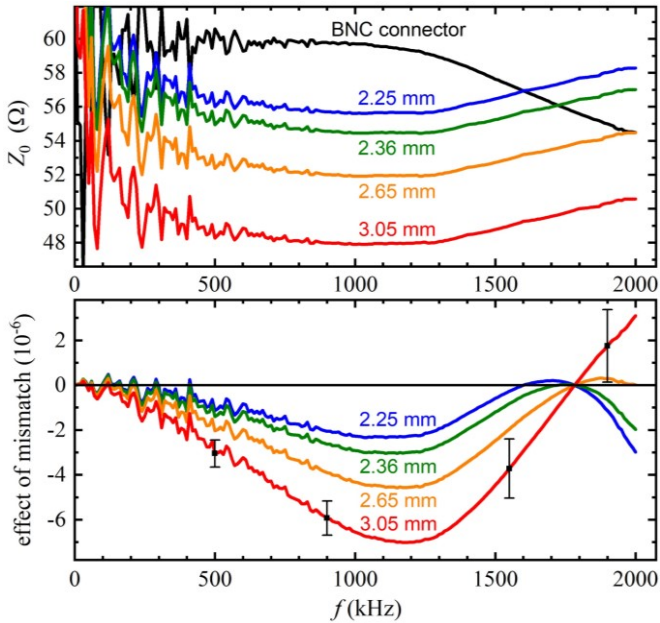

**Figure 4.** (Top) Characteristic impedance $Z_0$ of microstrip lines of different width as indicated on standard FR-4 printed circuit boards as well as the average characteristic impedance of four '50 Ω' BNC connectors. (Bottom) The effect of the impedance mismatch on the frequency dependence of the resistance standards. The exemplary error bars represent the uncertainty of the data.

We built two different types of cases for the resistance standards. At the type shown in figure 3, the BNC feedthroughs are mounted in line and with the same spacing of 22 mm as is standard at commercial LCR meters and impedance analysers. Therefore, the case can be directly connected to the BNC ports of the measuring instrument. This case is built from metal sheet and is not vacuum-tight.

A second type of case has been built that can be evacuated and, if desired, vented with dry nitrogen gas. This allows us to experimentally determine the contribution of surface layers of water, residual solvent, and grease to the frequency dependence of the resistance standards and to explore to which degree this can be reduced in high vacuum. A vacuum-tight case requires a higher wall thickness, an O-ring sealed flange, and two vacuum valves. As a result, the case is too heavy to be directly plugged into the LCR meter. Therefore, it stands in front of the LCR meter and is connected to it by short cables. Of course, also the non-vacuum-tight cases can be (and indeed were) temporarily placed opened in a vacuum chamber to remove residual solvent.

## 4. Experimental results

### 4.1 Auxiliary AC measurements at single SMD resistors

To allow the numerical calculation of the frequency dependence of the series-connected SMD resistors according to equation (1), the electrical AC properties of single SMD test pieces of each type were measured. The measuring instruments are a commercial capacitance bridge operated at 10 kHz and a precision LCR meter operated at 1 MHz. The capacitance in parallel to an SMD resistor could not be measured with the precision LCR meter because the ratio of reactive to active resistance is too large. Therefore, an impedance analyser at (10 to 110) MHz with a test fixture for SMD components was used. The measured value of the parallel capacitance is practically independent of frequency, which justifies using this value also at lower frequencies (with an appropriate uncertainty). This value is also consistent with the numerical simulation. All input quantities of the model equation (1) are summarized in the top part of table 1.

Concerning the dissipation factor of the parallel capacitance, $D_P$, a direct measurement is not possible because of the parallel thin-film resistance. To obtain at least an estimate, two SMD resistors of the same type were mounted in parallel at a small distance, first with the thin films facing each other and then with the thin films facing outwards. In both configurations, the capacitance between the two SMD resistors as well as the associated dissipation factor was measured at normal conditions with a commercial capacitance bridge. These measurements allow the estimation of the dissipation factor of the protective lacquer, which is assumed to dominate the dissipation factor of an SMD resistor. The dissipation factor of a bare ceramic substrate was also measured but is much smaller than that of the lacquer.

Finally, the results of the auxiliary measurements and the numerical simulations are used to quantitatively calculate the frequency dependence according to equation (1), as listed in the bottom part of table 1. Both the linear and the quadratic frequency dependence of the impedance standards are remarkably small. The results for four and twelve series-connected SMD resistors differ because of the different SMD size and the different number $N$ rather than the different manufacturer and fabrication technique. At 2 MHz, the total calculated AC-DC difference with the combined $k = 1$ uncertainty is $(-12.2 \pm 2.8)$ and $(-3.1 \pm 1.2)$ parts per million from the DC value, respectively, which is excellent.

### 4.2 Remarks on the dissipation factor

For the following investigation, the protective lacquer and the metal film have been removed from a single SMD test resistor. The remaining ceramic (probably $Al_2O_3$) with the metal caps was soldered in a coaxial metal case (with a shield between the high- and the low-side to minimize unwanted stray capacitances). The case can be evacuated by a turbomolecular pump with a dry-sealed booster pump. The ceramic was mechanically cleaned with a brush but had not been exposed to soldering flux or any solvent since its fabrication. The capacitance and the associated dissipation factor of the ceramic in various conditions has been measured by a commercial capacitance bridge in the frequency range up to 20 kHz. As shown in figure 5, the initial state had a large contribution not only from surface water layers (which are always present at normal conditions) but also from residual solvent that quickly desorbed in a high vacuum. Cleaning the whole object in ethanol in an ultrasonic bath and letting it dry for eight hours increased the dissipation factor due to residual ethanol, which afterwards quickly desorbed in vacuum.

**Table 1.** Electrical properties of the single thin-film SMD resistors (top) and the series-connected resistance standards (bottom). All uncertainties refer to coverage factor $k = 1$.

| Quantity | Susumu*, RG, size 0805, $N = 4$ | Vishay*, TNPU/TNPW, size 0603, $N = 12$ |
|---|---|---|
| rel. deviation from nominal resistance, 23 °C ($10^{-6}$) | 9.0 | -248.5 |
| temperature dependence of resistance (type, $10^{-6}$/°C) | linear, 5.6 ± 0.2 | parabolic, ≤ 0.5 at (23 to 26) °C |
| $C_P$ (fF) measured (simulated) | 83 ± 7 (115 ± 16) | 123 ± 7 (125 ± 7) |
| $D_P$ at 10 kHz ($10^{-3}$) | 1.0 ± 0.5 | 1.0 ± 0.5 |
| $C_C$ (fF) (incl. wire arc) | 35 ± 5 | 22 ± 5 |
| $D_C$ at 10 kHz ($10^{-3}$) | 0.24 ± 0.08 | 0.13 ± 0.04 |
| $C_{Cap}$ (fF) (incl. wire arc) | 14 ± 2 | 10 ± 2 |
| $C_R$ (fF) | 9 ± 3 | 7 ± 2 |
| $L_S$ (nH), simulation (strip approximation) | 3.0 (5.7) | 1.7 (1.7) |
| $L_S$ (nH), with wires | 7.2 ± 2.0 | 3.0 ± 1.0 |
| $C_{P,tot}$ (fF) model (directly measured) | -3 ± 3 (7.5 ± 15) | -43 ± 8 (-61 ± 15) |
| linear frequency term due to dissipative parallel capacitance ($10^{-6}$ MHz$^{-1}$) | -1.76 ± 0.41 | -0.89 ± 0.12 |
| linear freq. dependence due to dissipative case capacitances ($10^{-6}$ MHz$^{-1}$) | 0.46 ± 0.25 | 0.57 ± 0.35 |
| skin effect ($10^{-6}$ MHz$^{-1}$) (strip approximation) | 0.52 ± 0.21 (0.52 ± 0.21) | 0.46 ± 0.20 (0.46 ± 0.20) |
| total linear frequency dependence ($10^{-6}$ MHz$^{-1}$) | -0.78 ± 0.52 | 0.14 ± 0.42 |
| inductive quadratic freq. dependence ($10^{-6}$ MHz$^{-2}$) | 0.007 ± 0.003 | -0.082 ± 0.039 |
| resistive quadratic freq. dependence ($10^{-6}$ MHz$^{-2}$) | -2.68 ± 0.65 | -0.76 ± 0.22 |
| total quadratic frequency dependence ($10^{-6}$ MHz$^{-2}$) | -2.67 ± 0.65 | -0.84 ± 0.22 |

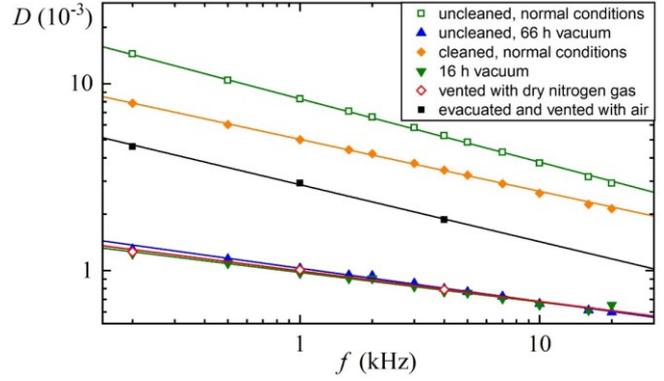

**Figure 5.** Dissipation factor of an SMD ceramic contacted at its metal caps and measured by a commercial capacitance bridge as a function of frequency. The ceramic has been treated in the sequence given in the legend. The solid lines are power-law least-square fits.

Remarkably, the changes in vacuum happen very quickly even before the turbomolecular pump in use reaches its full speed. Water and solvent molecules have a large permanent dipole moment, and at normal conditions, they cover every surface with several monolayers. The first few monolayers are very strongly bound to the surface and can only be removed by heating in ultra-high vacuum (which is not applied here). For the same reason, these dipole molecules cannot align themselves to an applied AC electrical field and, thus, do not cause dissipation. Because the molecules of the topmost layers are weakly bound, they can align themselves to an applied AC electrical field and cause dissipation, and for the same reason, they quickly desorb in vacuum. After the measurement in vacuum, the coaxial case was vented with dry nitrogen gas. Nitrogen is a non-polar molecule and its dissipation factor is several orders of magnitude smaller than that of water vapor and solvent at normal conditions. Therefore, as shown in figure 5, venting with dry nitrogen gas does not cause a significant increase of the dissipation factor.

In the frequency range up to 20 kHz, the measured dissipation factor decreases with frequency (as shown in figure 5). We have no equipment capable of measuring small dissipations factors of such small capacitances at higher frequencies, but the dissipation factor of polar molecules like water and ethanol above 20 kHz is expected to be roughly frequency-independent up to 10 MHz and then to increase toward a resonance in the microwave range. We do not know to which extent the polar molecules on the inner surfaces (for example, the water layer on the thin film below the protective lacquer) can be removed by cleaning and vacuum, but the results of figure 5 should at least describe the contribution of the rear side of a SMD resistor. Therefore, the values measured at the test pieces at 10 kHz are used as an estimate for the model calculation (see table 1). If the true dissipation factor in the whole frequency range up to 2 MHz would be smaller or up to two times larger than the estimate, this would not cause a substantial error of the calculated frequency dependence of the resistance standards. If the dissipation factor would be much larger (especially in the upper frequency range) this would not only cause a significant systematic error of the calculated frequency dependence but also a measurable effect of cleaning and vacuum on the frequency dependence of the standards, which in turn provides a valuable test opportunity.

*4.3 Series-connected thin-film resistors*

After mounting the printed circuit boards in the respective case and measuring the lead parameters of the microstrip lines, the SMD resistors were manually soldered to the respective printed circuit board under a long-distance low-magnification microscope with the help of a temporary Teflon underlay. Two sets of twelve series-connected Vishay* resistors named 12V1 and 12V2 and one set of four series-connected Susumu* resistors named 4S1 were mounted in non-vacuum-tight sheet cases which can be directly connected to the LCR meter. Photographs of two completed standards are shown in figure 6. Finally, another set of four series-connected Susumu* resistors named 4S2 was mounted into the vacuum-tight case.

Next, the deviation of the resistance standards from the nominal DC resistance value as well as the temperature dependence were measured by an 8½ digit DC multimeter (calibrated by a cryogenic current comparator [22]) and are listed in table 1. The temperature dependence is clearly smaller than the specifications given in section 3 (and also smaller than the $20 \times 10^{-6}$ per °C of common resistive wires used for Haddad [15] and Gibbings [16] resistors). Taking into account that the temperature in our laboratory is controlled within ±0.3 °C, the temperature dependence of the resistance standards is not a limitation.

Afterwards, the precision LCR meter was used to carry out frequency sweeps of the 4TP resistance standards. In the following, all results are plotted relative to the respective low-frequency value obtained from a least-square fit. Occasionally, the fitted low-frequency value was compared with the DC value measured by a cryogenic current comparator [22] and was found to agree within three parts per million.

At first, as shown in figure 7, a possible change of the frequency dependence of the resistance standards due to cleaning and placement in a high vacuum has been measured. A linear least-square fit to the data gives an upper limit for possible changes of $1.7 \times 10^{-6}$ MHz$^{-1}$ ($k = 1$) relative to the DC value, which is below the resolution of the LCR meter. This confirms that the dissipation factor of surface water and residual solvent in the whole frequency range is as small as expected. Consequently, it is not necessary to carry out the measurements in high vacuum or dry nitrogen, which greatly simplifies the application.

Finally, the three standards that can be directly connected to the LCR meter were measured in sequence and corrected for the lead parameters according to [18]. Figure 8 (top) presents the results of 12V1 and 12V2 relative to 4S1 acting as the reference standard. The measured and the calculated frequency dependences are in good agreement, even though the standards are very different. This demonstrates that the model calculations summarized in table 1 are reliable. The results obtained by two different LCR meters of the same model (the same as used in [18]) also agree with each other. Second-order polynomials fitted to the measured data differ from the calculated difference by at maximum $3.8 \times 10^{-6}$, which corresponds to half the resolution of the LCR meter at a frequency sweep. It cannot be much better than this.

The measurements also allow the determination of the frequency-dependent systematic error of the two LCR meters (figure 8 middle). Note that the systematic error of both LCR meters is different and shows some irregular structures with a magnitude of 50 to 70 parts per million that are eliminated from the measured frequency dependence of the resistance standards, as expected for a proper calibration. Finally, figure 8 (bottom) shows that the systematic error of the LCR meter combined with the total cable correction is well within the systematic uncertainty specified by the manufacturer. The average user of a precision LCR meter is well advised to use the manufacturer's specifications, but those willing to invest the effort described in this work and in [18] can reduce the uncertainty by a factor ranging from a hundred at lower frequencies to three hundred at 2 MHz.

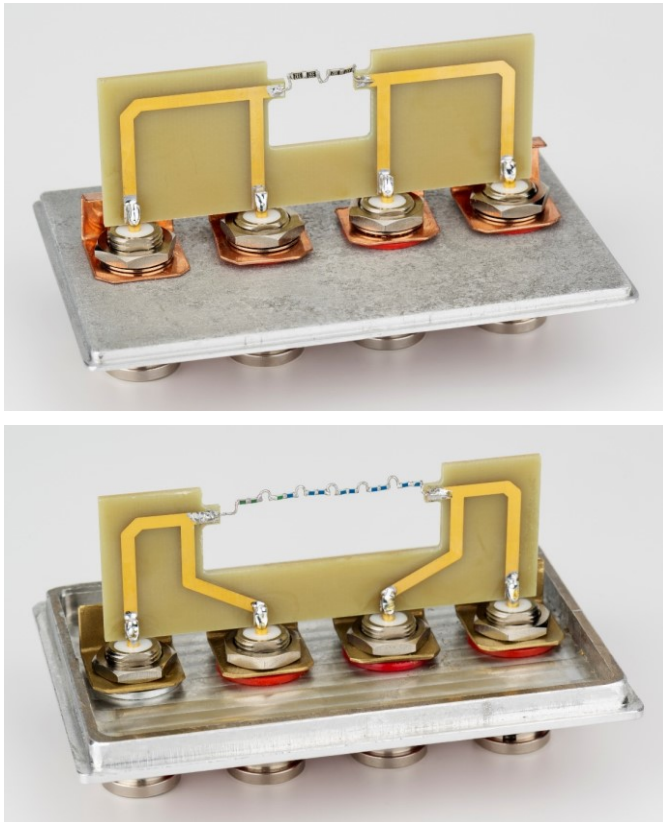

**Figure 6.** Photographs of the opened case with four series-connected Susumu* SMD resistors named 4S1 (top) and twelve series-connected Vishay* SMD resistors named 12V1 (bottom). The centre distance of two adjacent BNC connectors is 22 mm and can be used as a scale.

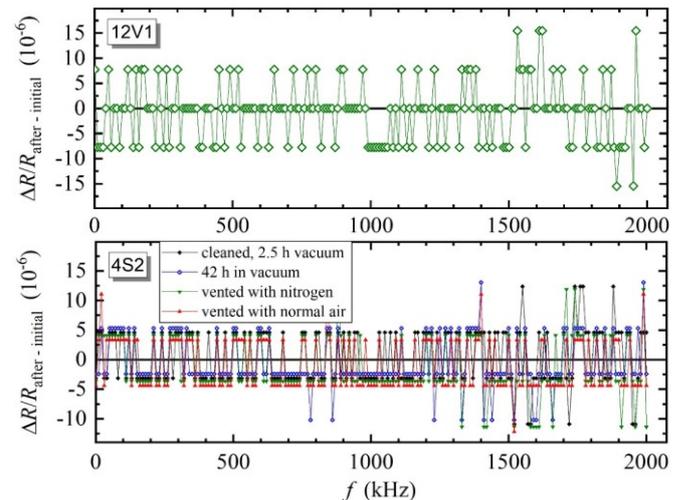

**Figure 7.** Change of the frequency dependence of the resistance standard 12V1 (top) measured by the LCR meter before and after cleaning and temporary placement in high vacuum (for two days with the case open and an end pressure of $3 \times 10^{-6}$ hPa). (Bottom) Change of the frequency dependence of the resistance standard 4S2 due to cleaning and high vacuum (end pressure $1 \times 10^{-6}$ hPa) and afterwards venting with either dry nitrogen gas or normal air.

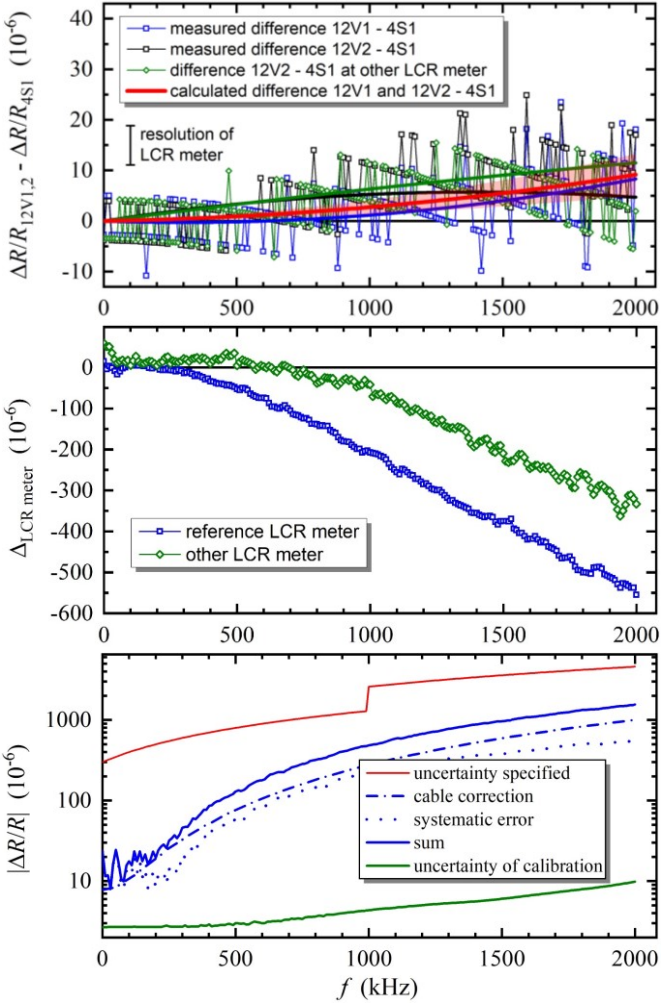

**Figure 8.** (Top) The two resistance standards 12V1 and 12V2 in terms of the reference standard 4S1 measured by two different LCR meters of the same model as a function of the frequency. The red solid line is the calculated difference according to the model and the light red band is the associated uncertainty ($k = 1$). The other solid lines are least-square fits of $2^{nd}$ order polynomials to the measured data. The bar indicates the resolution of the LCR meter at a frequency sweep. (Middle) The systematic error of the two LCR meters deduced from the measurements. (Bottom) The uncertainty specified by the manufacturer for the 10-k$\Omega$ range of the LCR meter, the sum of the cable correction according to [17] and the systematic error of the reference LCR meter (the blue data from the middle diagram), and the combined uncertainty of the calibration. The latter includes the uncertainty of the model calculation according to table 1, the effect of the impedance mismatch according to figure 4, the uncertainty of the cable correction, and the type-A uncertainty of the LCR meter.

## 5. Conclusion

It has been shown that resistance standards based on series-connected thin-film SMD resistors are independent of frequency within a few parts per million from the DC value and outperform all calculable wire resistors [15, 16] at frequencies up to a few megahertz. Such standards can be used to calibrate the respective resistance range of a precision LCR meter in the whole frequency range up to 2 MHz with a relative uncertainty of a few parts per million, which is considerably smaller than the systematic uncertainty specified by the manufacturer. This boost in precision requires a substantial effort but enables new applications of impedance metrology, such as graphene, 2D materials, and sensing of biomarkers [23–25] as well as high-frequency measurements of the quantum Hall resistance [26–28]. The basic concept presented in this work can be applied to a wide range of resistance values and may be of interest to national metrology institutes, calibration laboratories, and producers and users of precision LCR meters. Of course, these standards can also be used at even higher frequencies and at audio frequencies.

The idea behind this work is new and the development is not yet technically mature, so further improvements are possible: Firstly, SMD resistors with a smaller temperature coefficient and custom-made nominal values could be used. Secondly, the design of the printed circuit board could be slightly improved. And thirdly, to simplify the assembly, we will try a new design of printed circuit boards with separate cutouts for each SMD chip so that the soldering can be simplified.

Moreover, we have shown that a considerable contribution to the dissipation factor of dielectrics originates from those surface layers of water, residual solvent, and grease that quickly desorb in high vacuum. This finding might be relevant to the frequency dependence of capacitance standards in the audio-frequency range (which according to the Kramers-Kronig relation is directly related to the dissipation factor), to the AC-DC difference of capacitance standards [29], and to applications which are affected by dielectric relaxation. Finally, we have verified experimentally that these dissipation effects do not significantly contribute to the frequency dependence of the resistance standards.

## Acknowledgements

The authors would like to thank Luis Palafox for providing some thin-film SMD resistors, Oliver Kieler for providing another LCR meter of the same model, Dietmar Drung and Stephan Bauer for valuable discussions, and Niklas Abraham and Gerd Muchow for their highly valuable technical support.

**ORCID iDs**

Jürgen Schurr   https://orcid.org/0000-0001-7985-5770
Rolf H. Judaschke   https://orcid.org/0000-0002-0341-5194
Shakil A. Awan   https://orcid.org/0000-0002-8543-3328